\RequirePackage{amsthm}
\documentclass[sn-mathphys-num]{sn-jnl}%
\usepackage{lmodern}
\usepackage{graphicx}%
\usepackage{multirow}%
\usepackage{amsmath,amssymb,amsfonts}%
\usepackage{amsthm}%
\usepackage[title]{appendix}%
\usepackage{xcolor}%
\usepackage{textcomp}%
\usepackage{manyfoot}%
\usepackage{booktabs}%
\usepackage{algpseudocode}%
\usepackage{listings}%

\usepackage{lineno}
\usepackage{hyperref}
\usepackage{tcolorbox}
\usepackage[linesnumbered,ruled,vlined]{algorithm2e}
\usepackage{setspace}
\usepackage{float}

\theoremstyle{thmstyleone}%

\theoremstyle{thmstyletwo}%

\theoremstyle{thmstylethree}%

\begin{document}
\setstretch{1.5}
\title[Non-linear diffusion descriptions of memory-based dispersal]{Lattice-based stochastic models motivate non-linear diffusion descriptions of memory-based dispersal}

\author*[1]{\fnm{Yifei} \sur{Li}}\email{yifeili@hit.edu.cn}

\author[2, 3]{\fnm{Matthew} \sur{J Simpson}}

\author[1, 4]{\fnm{Chuncheng} \sur{Wang}}

\affil*[1]{\orgdiv{School of Mathematics}, \orgname{Harbin Institute of Technology}, \orgaddress{\city{Harbin}, \postcode{150001}, \country{P. R. China}}}

\affil[2]{\orgdiv{School of Mathematical Sciences}, \orgname{Queensland University of Technology (QUT)}, \orgaddress{\city{Brisbane}, \postcode{QLD 4001}, \country{Australia}}}

\affil[3]{\orgdiv{ARC Centre of Excellence for the Mathematical Analysis of Cellular Systems}, \orgname{QUT}, \orgaddress{\city{Brisbane}, \postcode{QLD 4001}, \country{Australia}}}

\affil[4]{\orgdiv{Zhengzhou Research Institute}, \orgname{Harbin Institute of Technology}, \orgaddress{\city{Zhengzhou}, \postcode{450046}, \country{P. R. China}}}

\abstract{The role of memory and cognition in the movement of individuals (e.g. animals) within a population, is thought to play an important role in population dispersal.  In response, there has been increasing interest in incorporating spatial memory effects into classical partial differential equation (PDE) models of animal dispersal.  However, the specific detail of the transport terms, such as diffusion and advection terms, that ought to be incorporated into PDE models to accurately reflect the memory effect remains unclear.  To bridge this gap, we propose a straightforward lattice-based model where the movement of individuals depends on both crowding effects and the historic distribution within the simulation.  The advantage of working with the individual-based model is that it is straightforward to propose and implement memory effects within the simulation in a way that is more biologically intuitive than simply proposing heuristic extensions of classical PDE models.  Through deriving the continuum limit description of our stochastic model, we obtain a novel nonlinear diffusion equation which encompasses memory-based diffusion terms. For the first time we reveal the relationship between memory-based diffusion and the individual-based movement mechanisms that depend upon memory effects. Through repeated stochastic simulation and numerical explorations of the mean-field PDE model, we show that the new PDE model accurately describes the expected behaviour of the stochastic model, and we also explore how memory effects impact population dispersal.}




\maketitle

\section{Introduction}
The effects of memory and cognition are thought to play an important role in several animal movement behaviours, such as home ranging, foraging and patrolling~\cite{FaganEL2013}.  Many different PDE models have been developed to study animal movement and dispersal~\cite{okubo2001diffusion,lam2022introduction}, with more recent attention being focused on the question of how to incorporate impacts of memory effects in population dynamics.  For example, Shi et al.~\cite{shi2020jdde} recently proposed the following delay diffusion model via a modified Fick’s law to describe the spatiotemporal distribution of a population density $U(x,t)> 0$,
\begin{equation}
\label{eq_shi}
\frac{\partial U(x,t)}{\partial t}=D_1\Delta U(x,t)+D_2\nabla \cdot(U(x,t)\nabla f(U(x,t-\tau))+g(U(x,t)).
\end{equation}
In this model the motion of individuals is modelled by combining linear diffusion, with diffusivity $D_1$, with a new nonlinear term $D_2\nabla \cdot(U\nabla f(U))$, which is often called \textit{memory-based diffusion} \cite{shi2020jdde,Song20196316,LIU20231bifurcation}. In this new term the function $f(U)$ encodes the dependence of memory-based diffusion on the memory of the density gradient at time $\tau$ before the present time.   The function $g(U)$ represents a standard source/sink term which can be used to model birth/death processes, and a standard choice would be to model logistic population growth by setting $g(U) = \lambda U (1 - U/K)$, where $\lambda > 0$ is the growth rate and $K > 0$ is the carrying capacity density.

The focus of this study is to examine the memory-based diffusion term in Equation~\eqref{eq_shi}. Very often this model is implemented by setting $f(U)=U$, and the physical motivation behind this choice of $f(U)$ remains unclear \cite{shi2020jdde,QiAN2020,Song2022consumer}. This kind of memory-based diffusion equation has been increasingly used as a foundation in various population models including both single species~\cite{Song20196316,Shi2019maturationdelay,QiAN2020,Wangyujia2022,Zhang2023advective,Xue2024,Shi2024dirichlet} and predator-prey models~\cite{Song2021hopf,Song2022consumer,Lishuai2023,LIU20231bifurcation,Wangyujia2023}. Many mathematical problems associated with equation \eqref{eq_shi} and its extensions, such as the uniform boundedness of solutions, are challenging due to the presence of memory delay in diffusion. 
 The recent review by Wang et al.~\cite{Wang2023open} summarises a range of open problems in memory-based PDE models at the present time.

A major challenge when working with continuum models alone is that complex individual-level behaviours are difficult to capture explicitly \cite{maini2004traveling,Jin2016JTB,yifei2022}. This challenge means that it is difficult to interpret the validity of existing continuum models in terms of \textit{how} these models represent memory. Heuristically, the memory-based diffusion term in equation \eqref{eq_shi} can be further modified to give a more general model
\begin{equation}
\label{eq_shi_modified}
\frac{\partial U(x,t)}{\partial t}=D_1\Delta U(x,t)+D_2\nabla \cdot(h(U(x,t))\nabla f(U(x,t-\tau))+g(U(x,t)),
\end{equation}
where $h(U)$ potentially depends on the joint effect of crowding and memory \cite{Wang2023open}. Implementing this generalisation could, in principle, allow this model to capture a broader range of dispersal phenomena than Equation~\eqref{eq_shi}. However, working at the continuum-level only does not make it clear how different choices of $h(U)$ relate to different behaviours of interest. 

One way of addressing this limitation of working with continuum models alone is to work with agent-based models, where a set of biologically motivated and interpretable \textit{rules} are directly encoded into a computational framework so that we can directly impose how individual agents (e.g. individual animals) behave in a certain situation~\cite{Couzin20021, Tang2010agent}. These kinds of individual-based models also have the advantage of introducing stochasticity into the simulations as the rule-based mechanisms are often implemented in a random walk framework, where the outcome of any particular realisation of the individual-based model can be different.  Incorporating stochasticity into a model of dispersal makes the outcome of the simulation model more realistic since biological and ecological experiments involve some level of stochasticity that is not captured by working with continuum models alone~\cite{Macfarlane2019}. While many lattice-based models have been proposed to explore how crowding effects impact the dispersal of individuals within populations~\cite{anguige2009one,simpson2010migration,yifei2022}, we are unaware of any previous study that has attempted to model spatial memory effect within a stochastic individual-based model.  Furthermore, in addition to proposing a simple memory-based dispersal mechanism in a discrete model, we can use coarse-graining techniques to explore how memory effects in the stochastic model relate to a population-level PDE description.  This process of coarse-graining will provide the first mechanistically-based derivation of a PDE describing the dispersion of a population of individuals undergoing dispersal with a memory effect.  A natural question that we will explore is how the new coarse-grained PDE model compares with previously-studied heuristic PDE models, such as Equation~\eqref{eq_shi}.

In this work, we build a discrete-to-continuous modelling framework based on an easy-to-interpret lattice-based model describing memory-based diffusion within a population of individuals. In Section~\ref{seU_dis}, we design a stochastic simulation algorithm where the movement mechanism of individuals depends on both a crowding effect modelled by an exclusion process, together with an effect based on the previous state of the system at an earlier time to represent a memory effect. The stochastic algorithm also includes a proliferation mechanism, and in this work we restrict the impact of the memory effect to the migration mechanisms only. In Section~\ref{seU_con}, we derive the continuum limit description of the stochastic model to obtain a novel PDE that approximately incorporates memory-based diffusion. As we point out, our derivation of the mean-field PDE leads to a novel memory function function~$h$.  This difference means that previous technical challenges in analysing the global boundedness of solutions of the new PDE can be addressed more easily than in standard heuristic approaches such as Equation \eqref{eq_shi}. In Section~\ref{seU_comp}, we illustrate how averaged data obtained from repeated stochastic simulations match with numerical solutions of the continuum limit PDE, and we explore how different choices in the discrete model affect the quality of the continuum-discrete match.  Furthermore, we also demonstrate how individuals' preference of moving towards memory gradient affects population dispersal. Inspired from the derivation of the memory-based diffusion term,
in Section~\ref{seU_reaction}, we further modify the reaction term in the continuum limit PDE model based on a lattice-based model with long-distance proliferation supported by experimental observations, which provides a novel perspective that the logistic-type reaction term in PDE models can be understood as a consequence of the volume-filling effect on population proliferation. 
In Section~\ref{seU_2D}, we extend our discrete-continuous modelling framework from a one-dimensional strip region to a two-dimensional plane, leading to abundant population dynamics associated with memory effects. Numerical algorithms for simulating the discrete model and solving the continuous model are available on \href{https://github.com/Yifei216/MemoryBased1}{GitHub}.

\section{Discrete model}
\label{seU_dis}
We consider a one-dimensional lattice with spacing $\Delta>0$. Each lattice site is indexed by $l$, and in any single realisation of the stochastic model each site is either vacant, $U_l=0$, or occupied by a single agent, $U_l=1$.  Discrete simulations are advanced through time using a simple fixed time step algorithm known as a random sequential update method~\cite{CHOWDHURY2005318}.  In any single time step of the discrete model, if there are $Q(t)$ agents on the lattice, we randomly select $Q(t)$ agents, one at a time, with replacement, and allow those agents to attempt to move.  The motility of individual agents is impacted by the state of the system at some previous time, and the direction of movement is unbiased so that the target site is chosen at random.  If the randomly-chosen target site is occupied then the potential motility event is aborted. Once $Q(t)$ times attempts are made, we again select $Q(t)$ agents, one at a time, with replacement, and allow those agents to attempt to proliferate.  If an agent proliferates, the daughter agent will be placed on a nearest neighbour lattice site and the target site is chosen at random.  Again, if the randomly-chosen target site is occupied then the potential proliferation event is aborted.  Once $Q(t)$ potential motility events and $Q(t)$ potential proliferation events have been attempted, time is advanced from $t$ to $t+\eta$ \cite{CHOWDHURY2005318}, and $Q(t + \eta)$ is updated accordingly.

\begin{figure}[t]
   \centering \includegraphics[width=0.6\textwidth]{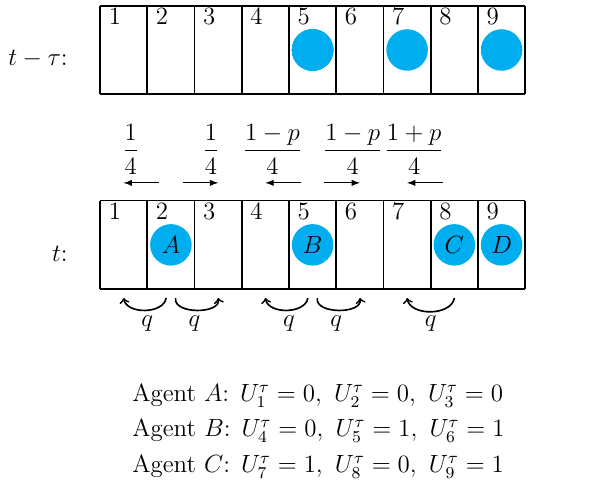}
   \caption{Schematic of the movement and proliferation mechanisms in the lattice-based model. Given the historic spatial distribution at $t-\tau$, the probabilities of agent $A$ at site 2 moving to either of the neighbouring vacant sites are $1/4$ per time step; the probabilities of agent $B$ at site 5 moving to either of the neighbouring vacant sites are $(1-p)/4$ per time step; the probability of agent $C$ at site 8 moving to site 7 is $(1+p)/4$ per time step, whereas agent $C$ at site 8 is unable to move to site $9$ owing to exclusion. The probability of agents proliferating per time step is always $q$.}
   \label{fig1}
\end{figure}

To model the time delay effect in the migration mechanism, we consider a relatively straightforward motility mechanism where the probability of movement for an agent at site $l$ depends on the historic occupancy of the sites at $l-1$, $l$ and $l+1$, see Figure~\ref{fig1}. More precisely, the possibility of attempting to move into site $l-1$ for the selected individual at $l$ is
\begin{equation}
    \nonumber
    P_{\text{left}}=
    \left\{
    \begin{aligned}
        &\frac{1-p}{4}, && \text{if}\quad U^\tau_{l}=1,\ U^\tau_{l-1}=0,\\
        &\quad \frac{1}{4}, && \text{if}\quad  U^\tau_{l}=0,\ U^\tau_{l-1}=0\ \text{or}\ U^\tau_{l}=1, U^\tau_{l-1}=1,\\
        &\frac{1+p}{4}&& \text{if}\quad  U^\tau_{l}=0,\ U^\tau_{l-1}=1,
    \end{aligned}
    \right.
\end{equation}
and the possibility of attempting to move into site $l+1$ is
\begin{equation}
    \nonumber
    P_{\text{right}}=
    \left\{
    \begin{aligned}
        &\frac{1-p}{4}, && \text{if}\quad U^\tau_{l}=1,\ U^\tau_{l+1}=0,\\
        &\quad \frac{1}{4}, && \text{if}\quad  U^\tau_{l}=0,\ U^\tau_{l+1}=0\ \text{or}\ U^\tau_{l}=1, U^\tau_{l+1}=1,\\
        &\frac{1+p}{4}&& \text{if}\quad  U^\tau_{l}=0,\ U^\tau_{l+1}=1,
    \end{aligned}
    \right.
\end{equation}
where $U^\tau_l$ represents the occupancy of site $l$ at time $t-\tau$ and parameter $p$ is the memory effect coefficient.   Regardless of the memory effect, any potential movement or proliferation event that would place an agent on an occupied lattice site is aborted owing to crowding effects.  This discrete mechanisms requires $\lvert p\rvert\le1$ so that $P_\text{left}+P_{\text{right}} \le 1$. No-flux boundary conditions are applied in the discrete model so that agents cannot step off the lattice at either boundary. An illustration of the movement and proliferation mechanisms in the discrete model is described in Figure~\ref{fig1}, and pseudo-code for a single realisation of the stochastic model is given in Algorithm~\ref{algorithm1}.

{
\begin{tcolorbox}
\begin{algorithm}[H]
\label{algorithm1}
\SetAlgoLined
 Create a one-dimensional lattice with $Q_0$ agents;\\
 Set $t=0$ and $Q(0)=Q_0$;\\
 \While{\normalfont $t<t_{\text{end}}$}{
 $t=t+\eta$,
 $Q(t)=Q(t-\eta)$,
 $B_1=0$;\\
 \While{$B_1<Q(t)$ }{
 $B_1=B_1+1$\;
 Randomly choose an agent;
   Calculate $P_{\text{left}}$ and $P_{\text{right}}$\;
   Draw a random variable: $\gamma_1\sim \textit{U}[0,1]$\;
    \uIf{\normalfont $\gamma_1<P_{\text{left}}$}{
       \If{\normalfont the left site is vacant}{
       Move the agent to its left site;
       }
       }
    \ElseIf{\normalfont $\gamma_1<P_{\text{left}}+P_{\text{right}}$}{
       \If{\normalfont the right site is vacant}{
       Move the agent to its right site;
       }
    }
  }
  Reset $B_1=0$\;
   \While{$B_1<Q(t)$ }{
 $B_1=B_1+1$; Randomly choose an agent\; Draw a random variable: $\gamma_1\sim \textit{U}[0,1]$\;
    \uIf{\normalfont $\gamma_1<q$}{
       \If{\normalfont the left site is vacant}{
       Place a new agent at its left site;
       }
       }
    \ElseIf{\normalfont $\gamma_1<2q$}{
       \If{\normalfont the right site is vacant}{
       Place a new agent at its right site;
       }
    }
  }
}
\caption{}
\end{algorithm}
\end{tcolorbox}
}

\section{Continuum limit description}
\label{seU_con}
By averaging the occupancy of site $l$ over $V$ identically prepared realisations of the stochastic model, we obtain the average occupancy at site $l$,
\begin{equation}
    \nonumber
    \bar{U}_l(t)=\frac{1}{V}\sum_{v=1}^VU^{(v)}_l(t),
\end{equation}
where $\bar{U}_l(t) \in [0,1]$ is the expected occupancy of site $l$ at time $t$ and $U^{(v)}_l(t)$ is the binary occupancy of site $l$ at time $t$ in the $v$-th identically prepared realisation of the discrete model. Subsequently, a discrete conservation statement for $\delta U_{l}$, which describes the change of the occupancy of a lattice site $l$ during a time step of duration $\eta$, can be developed as follows,
\begin{equation}
\label{conservationsform1}
\begin{aligned}
\delta U_{l}=&-\frac{1-p(\bar{U}^{\tau}_{l}-\bar{U}^{\tau}_{l-1})}{4}\bar{U}_{l}(1-\bar{U}_{l-1})-\frac{1+p(\bar{U}^{\tau}_{l+1}-\bar{U}^{\tau}_{l})}{4}\bar{U}_{l}(1-\bar{U}_{l+1})\\
&+\frac{1+p(\bar{U}^{\tau}_{l}-\bar{U}^{\tau}_{l-1})}{4}\bar{U}_{l-1}(1-\bar{U}_{l})+\frac{1-p(\bar{U}^{\tau}_{l+1}-\bar{U}^{\tau}_{l})}{4}\bar{U}_{l+1}(1-\bar{U}_{l})\\
&+q\bar{U}_{l+1}(1-\bar{U}_{l})+q\bar{U}_{l-1}(1-\bar{U}_{l}).
\end{aligned}
\end{equation}

The first two terms on the right-hand side of Equation~\eqref{conservationsform1} describe the expected decrease in occupancy of site $l$ owing to motility events out of site $l$, and similarly the next two terms represent the expected increase in occupancy of site $l$ due to motility events that would place agents at site $l$.  The final two terms on the right of Equation~\eqref{conservationsform1} describe the expected increase in occupancy of site $l$ owing to proliferation events. Note that this discrete conservation statement is approximate in the sense that each term on the right of Equation~\eqref{conservationsform1} involves products of terms that describe the probability of occupancy and the probability of vacancy of various lattice sites.  By simply interpreting these products as transition probabilities, we are implicitly assuming that the occupancy status of lattice sites is independent which amounts to ignoring correlation effects in the discrete model~\cite{johnston2012mean}.  These assumptions are known to impact the accuracy of the resulting continuum model for certain choices of parameters~\cite{Baker2010,Simpson2011corrected} and we will explore the impact of this standard mean-field approximation later when we compare averaged data from the discrete model with solutions of the resulting mean-field PDE.

To arrive at a continuous model we replace $\bar{U}_l(t)$ with a continuous function, $U(x,t)$, where $x = (l-1)\Delta$. Expanding terms involving $U(x \pm \Delta,t)$ about $x$ gives 
\begin{equation}
\label{eq_te1}
U(x\pm\Delta,t)=U(x,t)\pm\Delta\frac{\partial U(x,t)}{\partial x}+\frac{\Delta^{2}}{2}\frac{\partial^{2} U(x,t)}{\partial x^{2}}+\mathcal{O}\left(\Delta^{3}\right).
\end{equation}
Similarly, we replace $\bar{U}^\tau_{l\pm1}$ with $U(x,t-\tau)$, and expanding terms involving $U(x \pm \Delta,t-\tau)$ about $x$ gives
\begin{equation}
\label{eq_te2}
U(x\pm\Delta,t-\tau)=U(x,t-\tau)\pm\Delta\frac{\partial U(x,t-\tau)}{\partial x}+\frac{\Delta^{2}}{2}\frac{\partial^{2} U(x,t-\tau)}{\partial x^{2}}+\mathcal{O}\left(\Delta^{3}\right).
\end{equation}
For notational convenience, we rewrite \eqref{eq_te1} and \eqref{eq_te2} as
\begin{equation}
\nonumber
U_{\pm \Delta}=U\pm\Delta\frac{\partial U}{\partial x}+\frac{\Delta^{2}}{2}\frac{\partial^{2} U}{\partial x^{2}}+\mathcal{O}\left(\Delta^{3}\right),\quad
U_{\pm \Delta}^\tau=U^\tau\pm\Delta\frac{\partial U^\tau}{\partial x}+\frac{\Delta^{2}}{2}\frac{\partial^{2} U^\tau}{\partial x^{2}}+\mathcal{O}\left(\Delta^{3}\right).
\end{equation}
Substituting the truncated Taylor series representations of $U_{\pm\Delta}^\tau$ and $U_{\pm\Delta}$ into Equation~\eqref{conservationsform1}, dividing the resulting expression by $\eta$ and then considering the limit as $\Delta \to 0$ jointly as $\eta \to 0$ with the ratio $\Delta^2/\eta$ remaining finite leads to  
\begin{equation}
\label{eq_1}
\frac{\partial U}{\partial t}=D\frac{\partial}{\partial x}\left[ \frac{\partial U}{\partial x}-h\left(U\right)\frac{\partial U^\tau}{\partial x}\right]+f(U),
\end{equation}
where $D=\lim_{\Delta,\eta\to0}\Delta^2/4\eta$ is the diffusivity coefficient, and
\begin{equation}
\label{hf}
h(U)=2pU(1-U),\quad f(U)=rU(1-U).
\end{equation}
To obtain a well-defined continuum limit we require that all terms in Equation~\eqref{eq_1} be $\mathcal{O}(1)$.  Since we have $\Delta^2 / \eta \sim \mathcal{O}(1)$, we also require $r=\lim_{\eta \to 0} 2q/\eta \sim \mathcal{O}(1)$.  To achieve this we impose $q \sim \mathcal{O}(\eta)$, suggesting that the continuum limit model will be accurate if the time scale of cell migration is small compared to the time scale of proliferation.  This assumption turns out to be very reasonable since cells typically move on a time scale of minutes whereas the cell cycle is around 24 hours~\cite{Simpson2010cell}.

Intuitively, the flux of population depends on two factors: the current density gradient and the memory density gradient. Moreover, the influence of memory gradient relates to the function $h(U)$ which reflects the preference of individuals moving towards the historically vacant or occupied region. Recalling that the memory effect coefficient $p$ determines the possibilities of moving. When $p=0$, \eqref{eq_1} simplifies to the linear diffusion equation. When $p>0$, the memory effect leads to a positive $h(U)$, which decreases the population flux. In contrast, when $p<0$, the memory effect leads to a negative $h(U)$, which increases the population flux. Furthermore, recalling that $\lvert p \rvert\le1$ leads to that $\lvert h(U) \rvert\le1$, this model explicitly regulates the scale of the present gradient and the memory gradient. The form of $h(U)$ also shares some similarities with the chemotaxis model affected by the volume-filling effect, which is reasonable since the derivation of the volume-filling model is based on the assumption that cells carry a certain volume and cannot penetrate through each other \cite{Painter2002, Federica2020}.

We now demonstrate the global dynamics of solutions to \eqref{eq_1}. If Neumann boundary conditions are applied to \eqref{eq_1}, and the initial function $\phi(x,t)$, that is assumed to satisfy,
\begin{equation}\label{initphi}
\phi(x,t)\in C^{1,0}([0,L]\times[-\tau,0]),~~~~~~~\frac{\partial\phi}{\partial x}(x,t)|_{x=0,L}=0,~~~\text{for~}t\in[-\tau,0].
\end{equation}
is assigned, then we arrive at the following full model:
\begin{equation}\label{full}
\begin{aligned}
&\frac{\partial U}{\partial t}=D\frac{\partial}{\partial x}\left[ \frac{\partial U}{\partial x}-h\left(U\right)\frac{\partial U^\tau}{\partial x}\right]+f(U),~~~t>0,~~x\in(0,L),\\
&\frac{\partial U}{\partial x}\rvert_{x=0,L}=0,~~~t>0,\\
&U(x,t)=\phi(x,t),~~~t\in[-\tau,0],~x\in[0,L],
\end{aligned}
\end{equation}
where $f$ and $h$ are given in \eqref{hf}. The global boundedness of the weak solution to \eqref{full} for $x$ in a bounded domain $\Omega$ with smooth boundary is considered in \cite{Liu2024}, using the boundedness-by-entropy method that was initially proposed in \cite{Junger2015} for studying the boundedness of solutions for cross-diffusion systems. The main idea of the proof is to transform the equation \eqref{full} into an auxiliary equation by the derivative of the so-called entropy function, given by
$$
H(u)=u(\ln u-1)+(1-u)(\ln(1-u)-1)+H_1,
$$
with $H_1$ being a positive constant that $H(u)>0$ for $0<u<1$, and then, prove the global existence of solution to the auxiliary equation. The boundedness of the solution of \eqref{full} will be ensured by the boundedness of the inverse of derivative of entropy function. For $h$ and $f$ in \eqref{hf}, it is straightforward that
$$
f(u)H'(u)\leq C_f(1+h(u)),~~~~h(u)H''(u)=2p
$$
for $C_f:=\mathop{sup}\limits_{0<u<1}f(u)H'(u)$. Therefore, 
the assumption $\textbf{(H0)}$ in \cite{Liu2024} is satisfied, which will imply the boundedness of weak solution of \eqref{full} by Theorem 2.3 in \cite{Liu2024}, as long as $\phi\in(0,1)$ and satisfies \eqref{initphi}.
Furthermore, using Proposition 2.1 in \cite{shi2020jdde}, the weak solution turns to be classic, if higher smoothness is posed on the initial function $\phi$, that is, $\phi(x,t)\in C^{2+\alpha,0}([0,L]\times[-\tau,0])$ for some $\alpha\in(0,1)$.

Obviously, \eqref{full} admits a constant steady state $U_0\equiv 1$.
Let $U=W+U_0$. Then,
\begin{equation}\label{Wequ}
\frac{\partial W}{\partial t}=D\frac{\partial^2 W}{\partial x^2}+D\frac{\partial}{\partial x}\left(2pW(1+W)\frac{\partial W^\tau}{\partial x}\right)-rW(1+W).
\end{equation}
For ease of notation, we use $W_x$ to represent $\partial W / \partial x$. Let
$$
V(W)=\frac{1}{2}\int_0^LW_x^2(t,x)dx+\frac{D}{2}\int_0^L\int_{t-\tau}^tW_{xx}^2(s,x)dsdx.
$$
It then follows that
$$
\begin{aligned}
\frac{dV(W)}{dt}&=D\int_0^L  W_x[W_x+2pW(1+W)W^\tau_x]_{xx}
dx-\int_0^LW_x[rW(1+W)]_xdx\\
&\quad+\frac{D}{2}\int_0^LW_{xx}^2dx-\frac{D}{2}\int_0^L(W^\tau_{xx})^2dx-r\int_0^L(1+2W)W_x^2dx\\
&\leq-\frac{D}{2}\int_0^LW_{xx}^2dx-\frac{D}{2}\int_0^L(W^\tau_{xx})^2dx+2pD(\|W\|+\|W\|^2)\int_0^L|W_{xx}W^\tau_{xx}|dx\\
&\quad-r\int_0^L(1+2W)W_x^2dx\\
&\leq0\
\end{aligned}
$$
as long as $W$ is in the small neighbourhood of zero solution.
Therefore, the zero solution of \eqref{Wequ} is asymptotically stable, which further implies the asymptotic stability of $U_0\equiv 1$ for \eqref{full}.

We remark that the global boundedness of solutions closely relates to the form of the memory-based diffusion term. Intuitively, 
$\lvert h(U) \rvert\le1$ for $U\in[0,1]$ (caused by $\lvert p \rvert\le1$), suggests that the influence of the historic distribution on the migration of individuals is limited.
If $h(U)=\tilde{D}U$ for some $\tilde{D}\in\mathbb{R}$, which goes back to the previous setting in \cite{shi2020jdde}, the uniform boundedness of solutions still remains open. In this case, it also has been shown in \cite{shi2020jdde} that the local stability of positive constant steady state $U_0\equiv 1$ is completely determined by the value of $|\tilde{D}|$, but independent of the time delay. However, if $h(U)$ is given by \eqref{hf}, the positive steady state $U_0$ is always asymptotically stable, as the linearisation of \eqref{eq_1} is given by ${\partial U}/{\partial t}=D{\partial^2 U}/{\partial x^2}-rU$, which is independent of $h$.
Furthermore, for such $h$ in~\eqref{hf}, it can be proved that there also exists Hopf bifurcation for~\eqref{eq_1}, if another maturation time delay $\sigma$ is posed in reaction term, that is, $f(U)=rU(1-U(t-\sigma))$, while the bifurcated periodic solution must be spatially homogeneous. This result is also very different from the one in \cite{Shi2019maturationdelay} for the case $h(U)=\tilde{D}U$, in that the bifurcated periodic solution can be either spatially homogeneous or spatially inhomogeneous, depending on the values of $|\tilde{D}|$.

\section{Comparisons between continuous and discrete models}
\label{seU_comp}
For the lattice-based model, we consider a one-dimensional lattice with 200 sites and 40 agents, where the agents are initially located in the middle of the region. Performing 1000 identically-prepared individual realisations, we average the stochastic data to obtain the discrete results of population densities, denoted as $\bar{U}$. For these results we consider $\Delta=1$ leading to a continuous model defined on the region $x\in[0,200]$. For the numerical solutions of \eqref{eq_1}, we use the method of lines. That is, we discretise \eqref{eq_1} in space and numerically solve the corresponding ordinary differential equations system. The discretisation form and numerical algorithm are given in the appendix. 

\begin{figure}[t]
   \centering \includegraphics[width=0.95\textwidth]{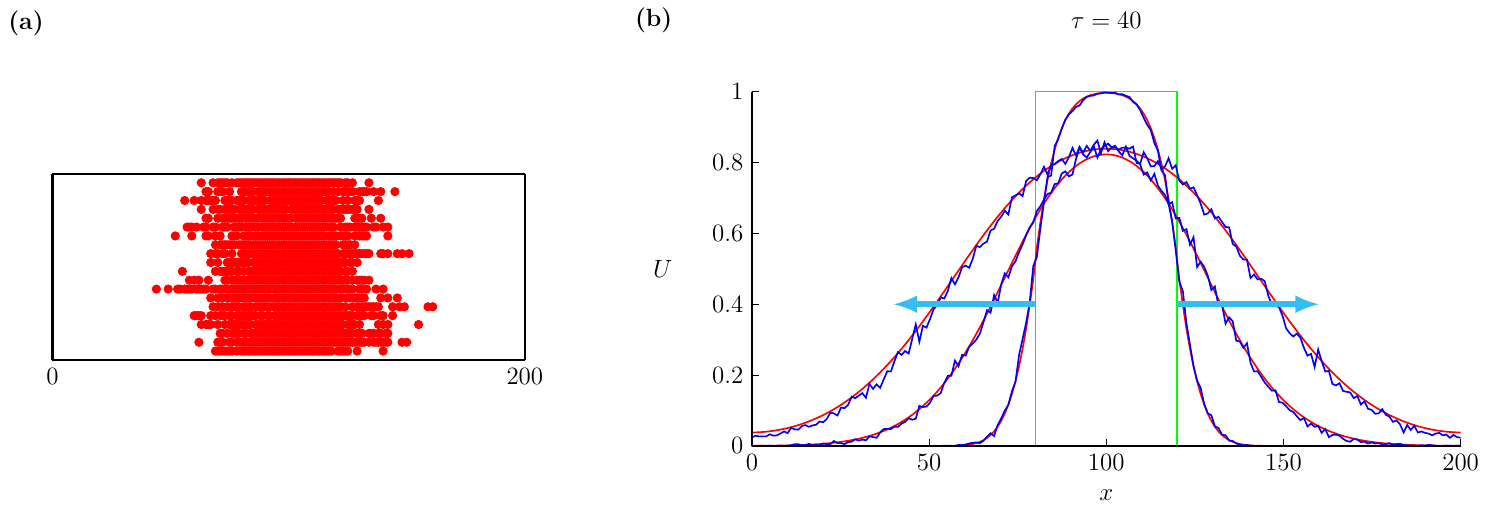}
   \caption{Panel (a) shows a snapshot of 20 times of stochastic realisations at $t=300$. Panel (b) overlaps the numerical solutions of \eqref{eq_1} (red curves) and the averaged stochastic results (blue curves). The green line indicates the initial distribution. We set $q=1\times10^{-3}$, $p=0.5$, $\tau=40$, $\Delta=1$ and $\eta=1$, and show the results at $t=100$, $t=1000$ and $t=2000$.}
   \label{fig2}
\end{figure}

The dispersal profiles of the population obtained by the continuous and discrete models are presented in Figure~\ref{fig2}. We place 40 agents in the middle of the region, which corresponds to the initial distribution $U(x,0)=1$ for $x\in[40,60]$ and $U(x,0)=0$ elsewhere. The historic distribution for $t\in(-\tau,0)$ is set to be the same as the initial distribution at $t=0$. Unless otherwise stated, we will always consider such historic distribution, which is the same as the initial distribution at time $t=0$. We then set $q=1\times10^{-3}$, $p=0.5$, $\tau=40$, $\Delta=1$ and $\eta=1$, and generate the discrete and continuous results at $t=100$, $t=1000$ and $t=2000$. As indicated by the arrows, the population disperses through migration and proliferation, and will eventually fill the whole space. Intuitively, the continuum and discrete results reasonably match with small errors.

\begin{figure}[H]
   \centering \includegraphics[width=0.8\textwidth]{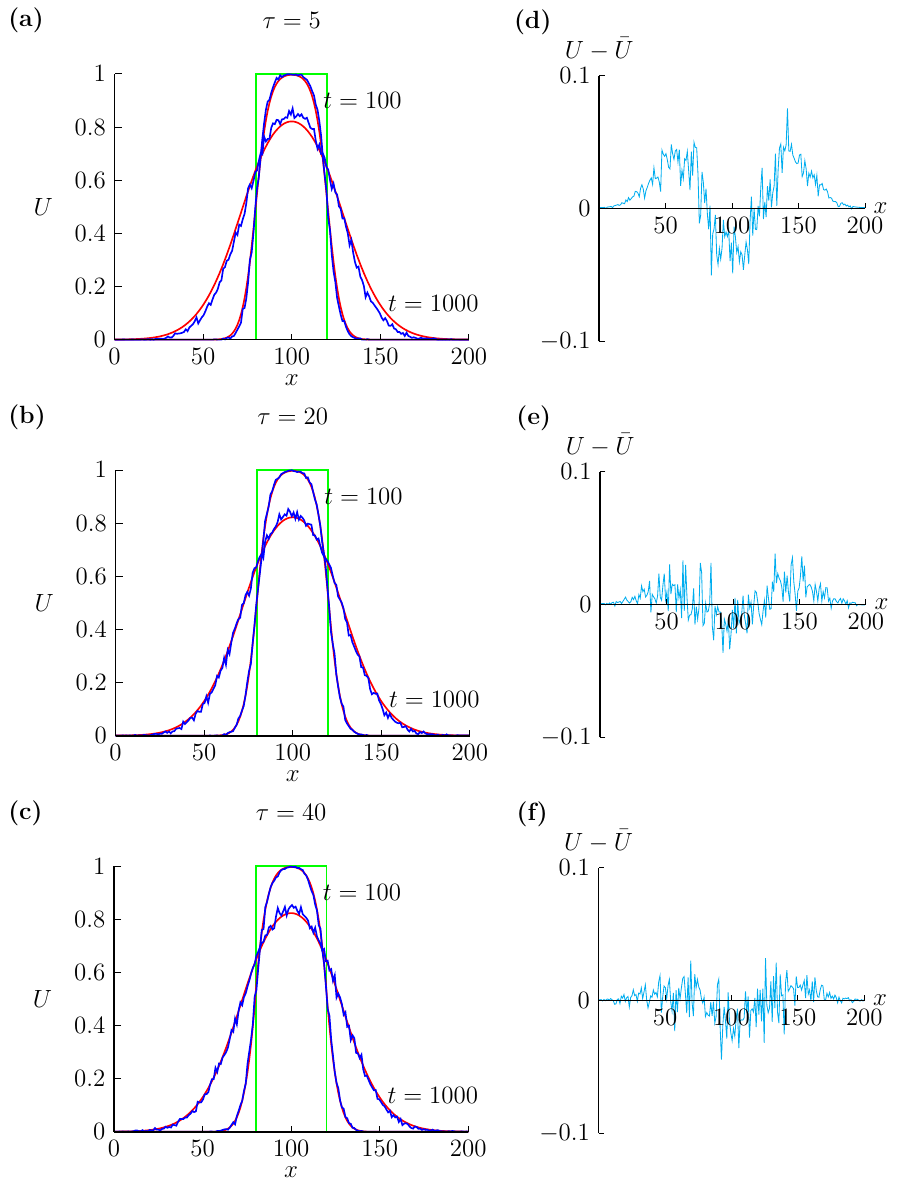}
   \caption{Comparisons of the numerical solutions of \eqref{eq_1} and the averaged stochastic results. We set $q=1\times10^{-3}$, $p=0.5$, $\Delta=1$ and $\eta=1$. Moreover, $\tau=5$ in (a), $\tau=20$ in (b) and $\tau=40$ in (c). The difference between the numerical solutions of \eqref{eq_1}, $U$, and the averaged stochastic results, $\bar{U}$, are given in (d)--(f) corresponding to different values of $\tau$.}
   \label{fig3}
\end{figure}

A simple visual comparison of the averaged discrete data and the solution of the continuum model in Figure \ref{fig2} indicates that the solution of the continuum limit PDE model provides a reasonably accurate way of predicting the evolution of the average density from the discrete simulations.  Despite this, the continuum model is still an approximation and discrepancies between the solution of the continuum limit PDE model and repeated simulations of the stochastic model can arise for several reasons. Firstly, the stochasticity of discrete simulations can lead to fluctuations in our density estimate, while this impact can be very small since we have averaged a large number of single-prepared realisations. Secondly, in deriving the continuum limit we require the time scale of proliferation to be much smaller than the time scale of migration to obtain a well-defined continuum limit. Thirdly, the assumption that the occupancy status of lattices is an approximation that can introduce discrepancies.   Despite these various approximations, the solution of the PDE model can still provide a useful approximation that we can further demonstrate by generating additional dispersal profiles with different choices of $\tau$ in Figure~\ref{fig3}(a)--(c).  Here we plot a measure of the discrepancy between the continuum and averaged stochastic density at $t=800$ in Figure~\ref{fig3}(d)--(f), where we see that the discrepancies are small and we see that discrepancy decreases as $\tau$ increases. This is because the current distribution at time $t$ and the historic distribution at time $t-\tau$ are less correlated as $\tau$ increases.

\begin{figure}[H]
   \centering \includegraphics[width=0.95\textwidth]{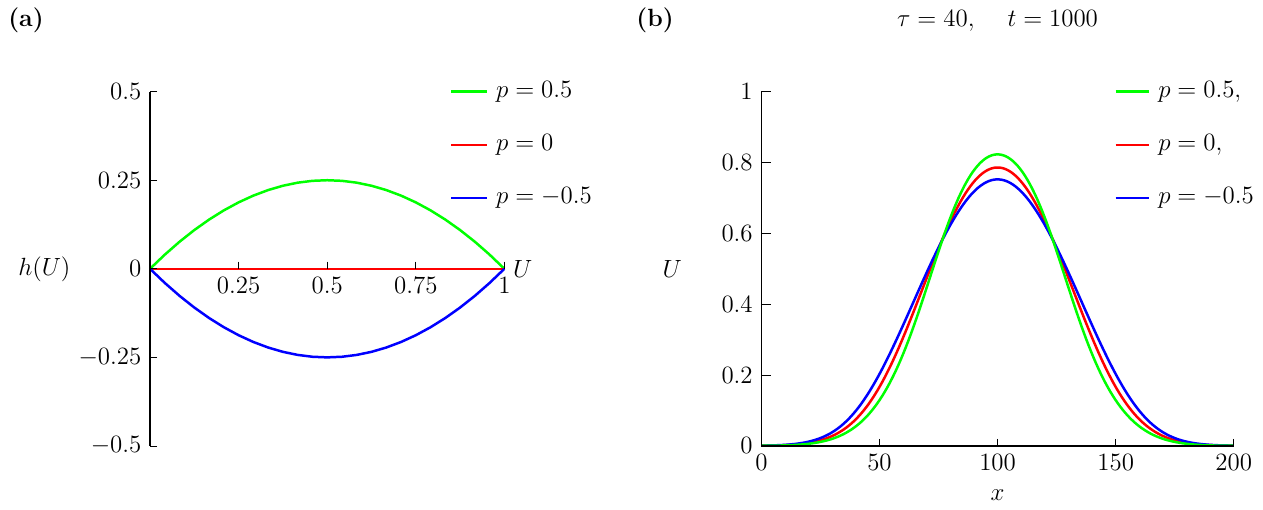}
   \caption{Comparisons of the dispersal profiles with different choices of $p$. Panel (a) shows $h(U)=2pU(1-U)$ with $p=0.5$ (green), $p=0$ (red) and $p=-0.5$ (blue), respectively. Panel (b) shows the numerical solutions of \eqref{eq_1} at time $t=1000$ with $q=1\times10^{-3}, \Delta=1, \eta=1, \tau=40$, where we consider $p=0.5$ (green), $p=0$ (red) and $p=-0.5$ (blue), respectively. }
   \label{fig4}
\end{figure}

To investigate how the memory effect influences population dispersal, we present additional dispersal profiles of the population with different choices of $p$ in Figure~\ref{fig4}. We first consider $p=0$, which represents that the movement of individuals is not influenced by the historic distribution. In this case, the movement mechanism of individuals corresponds to a simple exclusion-based random walk process and the memory-based diffusion simplifies to the linear diffusion equation. Then we consider a positive $p$, suggesting a movement mechanism where individuals prefer to move into the region that used to be occupied by others. In contrast, a negative $p$ suggests a movement mechanism where individuals prefer to move into the region that used to be vacant. Note that such preference is based on the density gradient at $\tau$ time ago. If an agent is located at a site where the site and its neighbouring sites are all vacant or occupied, the memory effect will not influence the probability of moving.

From the results presented in Figure~\ref{fig4}, the memory effect intuitively promotes population dispersal for $p > 0$, while reducing dispersal for $p < 0$. Although death events are not considered in the current model, it is expected that memory-based diffusion would influence the extinction of populations, since the extinction or survival of bistable populations is influenced by the dispersal process~\cite{yifei2022}.

\section{Modified source term}
\label{seU_reaction}
\begin{figure}[H]
   \centering \includegraphics[width=0.95\textwidth]{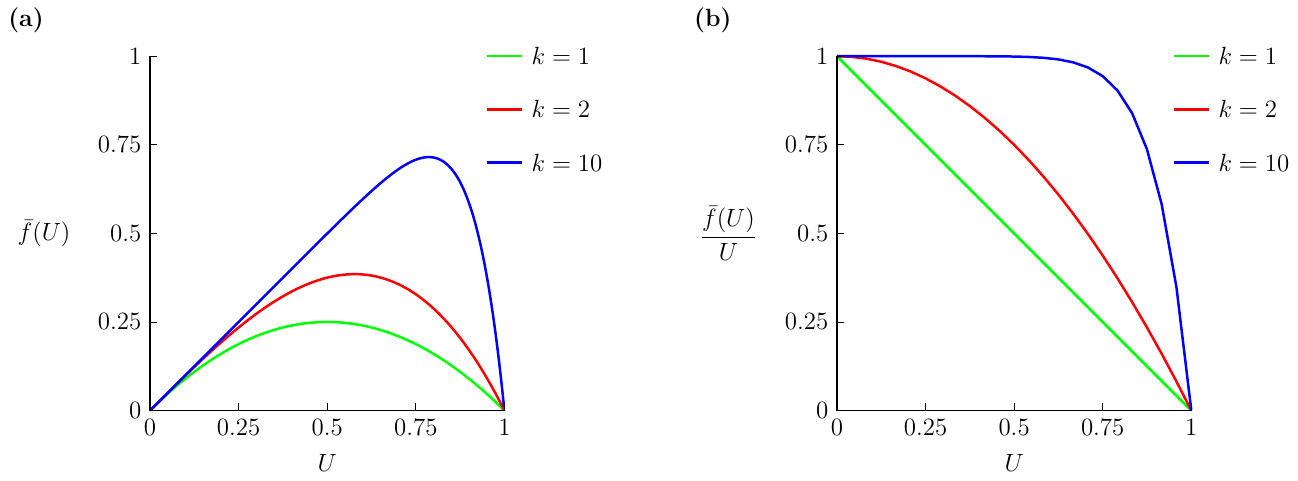}
   \caption{Comparisons of $\bar{f}(U)$ and $\bar{f}(U)/U$ with $r=1$ and different choices of $k$. }
   \label{fig5}
\end{figure}

It is interesting that $h(U)$ and $f(U)$ share similar forms. As the proliferation mechanism in the stochastic event is an exclusion process, where the newborn agents cannot penetrate through other agents, the derivation of the reaction term could also be interpreted as a consequence of the volume-filling effect. This perspective is quite different from the classical interpretation based on the ordinary differential equation models, where $g(U)=\lambda U(1-U)$ represents the logistic growth with limited resources and a normalised carrying capacity \cite{murray2002mathematical}.  However, the volume-filling effect may not play the same role in the movement and proliferation events of individuals. Based on an experimental observation, where the time-lapse movies of cell proliferation show that daughter cells are often generated some distance from the mother
cell~\cite{Druckenbrod2005}, we now consider a new proliferation mechanism of individuals in the stochastic model. Instead of assuming that the selected agent attempts to place a daughter agent into its nearest neighbouring sites, we assume that the selected agent is able to place the daughter agent within a certain distance on the lattice. More specifically, we allow the agent at site $l$ to attempt to place a daughter agent at most $k$ lattice sites away. This leads to the conservation statement given below:
\begin{equation}
\nonumber
\begin{aligned}
\delta U_{l}=&-\frac{1-p(\bar{U}^{\tau}_{l}-\bar{U}^{\tau}_{l-1})}{4}\bar{U}_{l}(1-\bar{U}_{l-1})-\frac{1+p(\bar{U}^{\tau}_{l+1}-\bar{U}^{\tau}_{l})}{4}\bar{U}_{l}(1-\bar{U}_{l+1})\\
&+\frac{1+p(\bar{U}^{\tau}_{l}-\bar{U}^{\tau}_{l-1})}{4}\bar{U}_{l-1}(1-\bar{U}_{l})+\frac{1-p(\bar{U}^{\tau}_{l+1}-\bar{U}^{\tau}_{l})}{4}\bar{U}_{l+1}(1-\bar{U}_{l})\\
&+q(\bar{U}_{l+1}+\bar{U}_{l+1}\bar{U}_{l+2}+...+\prod\limits_{i=1}^k\bar{U}_{l+i})(1-\bar{U}_{l})\\
&+q(\bar{U}_{l-1}+\bar{U}_{l-1}\bar{U}_{l-2}+...+\prod\limits_{i=1}^k\bar{U}_{l-i})(1-\bar{U}_{l}).
\end{aligned}
\end{equation}
If $l+i>N$ or $l-i<1$ with $i=1,...,k$, we transform $l+i$ into $l+i-N$ or transform $l-i$ into $N+l-i$ due to the application of periodic boundary condition, which leads to the same population dynamics as with the no-flux boundary conditions when we consider a symmetric initial distribution. Following the derivation steps as in deriving \eqref{eq_1}, we obtain  
\begin{equation}
\label{eq_2}
\frac{\partial U}{\partial t}=D\frac{\partial}{\partial x}\left[ \frac{\partial U}{\partial x}-h\left(U\right)\frac{\partial U^\tau}{\partial x}\right]+\bar{f}(U),
\end{equation}
where
\begin{equation}
\label{f2}
\bar{f}(U)=rU(1-U^k),
\end{equation}
see Figure~\ref{fig5} for a schematic illustration of $\bar{f}(U)$ and the corresponding per capita growth rates with several choices of $k$. Intuitively, long-distance proliferation enhances the growth rates of populations, especially the growth rates at high densities. Note that such long-distance proliferation does not result in the weak Allee effect, since the per capita growth rate does not decline in low densities~\cite{Hastings2005}.   This kind of growth law given by \eqref{f2} is sometimes known as the Richards' equation~\cite{Richards1959}, and this model has been used to model various types of biological population growth dynamics~\cite{Simpson2022}. 

\begin{figure}[H]
   \centering \includegraphics[width=\textwidth]{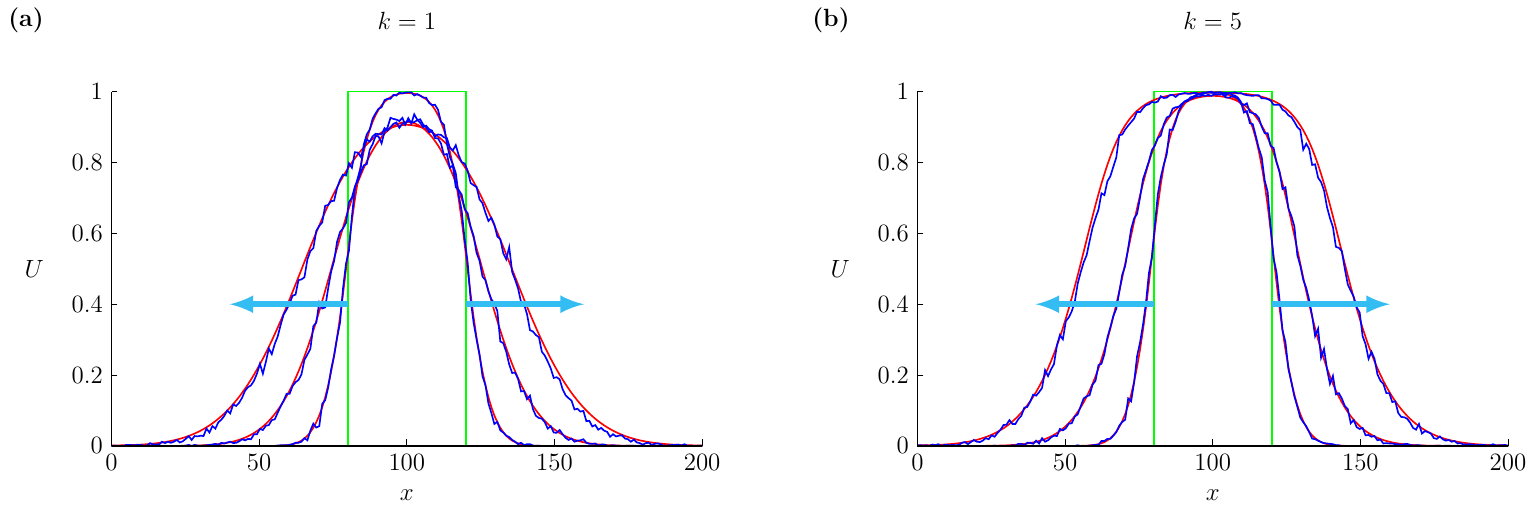}
   \caption{The numerical solutions of \eqref{eq_2} (red curves) and the averaged stochastic results (blue curves) with $k=1$ and $k=5$. The green line indicates the initial distribution. We set $q=1\times10^{-3}$, $p=0.5$, $\tau=40$, $\Delta=1$ and $\eta=1$, and show the results at $t=100$, $t=500$ and $t=1000$.}
   \label{fig6}
\end{figure}

The comparisons between the numerical solutions of \eqref{eq_2} with the averaged stochastic results with $k=1$ and $k=5$ are given in Figure~\ref{fig6}, which present a good match. Although we do not provide other results, an observation is that the errors increase when $k$ becomes larger. This is reasonable since a larger $k$ corresponds to the stochastic simulations with more successful attempts of birth events, while we need the time scale of birth events to be much smaller than the time scale of movement events. Moreover, the correlation effect becomes more important when we evaluate the occupancy of each site with the information of more sites. 

Results in Figure~\ref{fig6} suggest that the long-distance proliferation shaped the moving front of individuals. Although the monostable structure, where the reaction term is positive for $U\in(0,1)$, always leads to the survival of populations, the long-distance proliferation seems to drive the population to form a constant profile within a shorter time. This may dramatically influence the survival or extinction of populations if the population dynamics possess a bistable structure where the reaction term becomes negative in low densities. The interplay between memory effect and long-distance proliferation may lead to abundant dynamical behaviours. However, introducing a bistable reaction term into our discrete-continuous modelling framework usually needs a careful design of the growth mechanisms of individuals associated with complicated crowding effects \cite{yifei2022}. We will not delve into this problem in this manuscript, but leave it for future work.

\begin{figure}[H]
   \centering \includegraphics[width=0.95\textwidth]{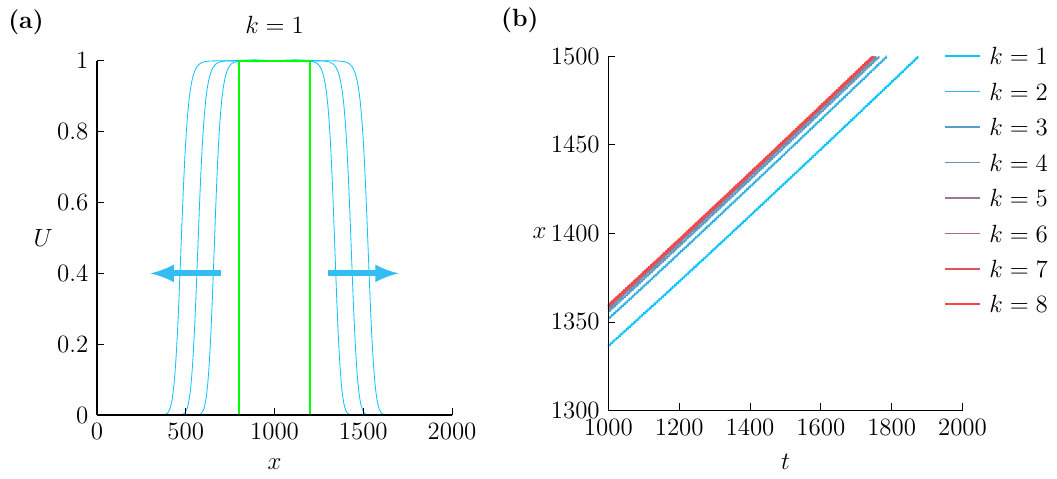}
   \caption{Panel (a) shows the numerical solutions of \eqref{eq_2} with $k=1$. Panel (b) shows the position of the moving fronts with $k=1,2,...,8$. We set $D=1$, $r=0.01$, $\tau=5$ and consider the initial distribution $U(x,0)=1$ for $x\in[800,1200]$ and $U(x,0)=0$ elsewhere.}
   \label{fig7}
\end{figure}

Another insight is that, although the different ranges in which a new daughter agent can be placed lead to different nonlinear terms, the long-distance proliferation seems only to shape the spreading profile and has no contribution to speeding up or slowing down the population propagation. As shown in Figure~\ref{fig7}, we present the propagation profiles in a very large domain in Figure~\ref{fig7}(a) so that the moving fronts are away from boundaries for a quite long time. We further show how the position of the wave changes in time with $k=1,2,...,8$ in Figure~\ref{fig7}(b), where the position of the wave is marked by $U=0.5$. With different $k$, the population generates the same propagation speed, which means that the essential effect of the growth mechanism on population propagation comes from the linear term. This is consistent with the intuition from analytical studies of reaction-diffusion equations with logistic-type reaction terms, where $\bar{f}(U)\le U\bar{f}'(0)$ implies that the spreading dynamics are linearly determined~\cite{Hadeler1975, carlos2013}. Moreover, further modifications and analysis need to be made in memory-based models, so that the discrete simulations and continuous models can provide a more comprehensive explanation for the interplay between nonlinear diffusion and growth mechanisms. Within the scope of our knowledge, the spreading dynamics of the memory-based diffusion model have not been investigated.

\section{Extend to a two-dimensional model}
\label{seU_2D}
Compared to a one-dimensional lattice, a two-dimensional plane is appropriate for modelling a wider range of population dynamics~\cite{Martinson2020}. However, it is challenging to incorporate the influence of memory effect on the movement of individuals in different directions. Now we propose a two-dimensional model with a simple rule distinguishing the memory effect in horizontal and vertical directions. For the two-dimensional model, the movement of a selected agent with the coordinate $(i, j)$, where $i, j=1,2,3,...$, depends on the historic occupancy of the sites $(i, j)$, $(i-1, j)$, $(i+1, j)$, $(i, j-1)$, $(i, j+1)$. The possibility of attempting to move to the left or right depends on the historic occupancy of the sites $(i,j)$, $(i-1, j)$ and $(i+1, j)$, and the possibility of attempting to move upwards or downwards depends on the historic occupancy of the sites $(i,j)$, $(i, j-1)$ and $(i, j+1)$. For attempting to move into site $(i-1,j)$, the possibility is
\begin{equation}
    \nonumber
    P_{\text{left}}=
    \left\{
    \begin{aligned}
        &\frac{1-p_h}{8}, && \text{if}\quad U^\tau_{i,j}=1,\ U^\tau_{i-1,j}=0,\\
        &\quad \frac{1}{8}, && \text{if}\quad  U^\tau_{i,j}=0,\ U^\tau_{i-1,j}=0\ \text{or}\ U^\tau_{i,j}=1, U^\tau_{i-1,j}=1,\\
        &\frac{1+p_h}{8}&& \text{if}\quad  U^\tau_{i,j}=0,\ U^\tau_{i-1,j}=1,
    \end{aligned}
    \right.
\end{equation}
For attempting to move into site $(i+1,j)$, the possibility is
\begin{equation}
    \nonumber
    P_{\text{right}}=
    \left\{
    \begin{aligned}
        &\frac{1-p_h}{8}, && \text{if}\quad U^\tau_{i,j}=1,\ U^\tau_{i+1,j}=0,\\
        &\quad \frac{1}{8}, && \text{if}\quad  U^\tau_{i}=0,\ U^\tau_{i+1,j}=0\ \text{or}\ U^\tau_{i,j}=1, U^\tau_{i+1,j}=1,\\
        &\frac{1+p_h}{8}&& \text{if}\quad  U^\tau_{i,j}=0,\ U^\tau_{i+1,j}=1,
    \end{aligned}
    \right.
\end{equation}
For attempting to move into site $(i,j-1)$, the possibility is
\begin{equation}
    \nonumber
    P_{\text{up}}=
    \left\{
    \begin{aligned}
        &\frac{1-p_v}{8}, && \text{if}\quad U^\tau_{i,j}=1,\ U^\tau_{i,j-1}=0,\\
        &\quad \frac{1}{8}, && \text{if}\quad  U^\tau_{i,j}=0,\ U^\tau_{i,j-1}=0\ \text{or}\ U^\tau_{i,j}=1, U^\tau_{i,j-1}=1,\\
        &\frac{1+p_v}{8}&& \text{if}\quad  U^\tau_{i,j}=0,\ U^\tau_{i,j-1}=1,
    \end{aligned}
    \right.
\end{equation}
where $p_h\in[-1,1]$ and $p_v\in[-1,1]$ represent the influence of memory effect on the migration of population in the horizontal and vertical directions, respectively. For attempting to move into site $(i,j+1)$, the possibility is
\begin{equation}
    \nonumber
    P_{\text{down}}=
    \left\{
    \begin{aligned}
        &\frac{1-p_v}{8}, && \text{if}\quad U^\tau_{i,j}=1,\ U^\tau_{i,j+1}=0,\\
        &\quad \frac{1}{8}, && \text{if}\quad  U^\tau_{i}=0,\ U^\tau_{i,j+1}=0\ \text{or}\ U^\tau_{i,j}=1, U^\tau_{i,j+1}=1,\\
        &\frac{1+p_v}{8}&& \text{if}\quad  U^\tau_{i,j}=0,\ U^\tau_{i,j+1}=1,
    \end{aligned}
    \right.
\end{equation}
The corresponding conservation statement is
\begin{equation}
\nonumber
\begin{aligned}
\delta U_{i,j}=&-\frac{1-p_h(\bar{U}^{\tau}_{i,j}-\bar{U}^{\tau}_{i-1,j})}{8}\bar{U}_{i,j}(1-\bar{U}_{i-1,j})-\frac{1+p_h(\bar{U}^{\tau}_{i+1,j}-\bar{U}^{\tau}_{i,j})}{8}\bar{U}_{i,j}(1-\bar{U}_{i+1,j})\\
&+\frac{1+p_h(\bar{U}^{\tau}_{i,j}-\bar{U}^{\tau}_{i-1,j})}{8}\bar{U}_{i-1,j}(1-\bar{U}_{i,j})+\frac{1-p_h(\bar{U}^{\tau}_{i+1,j}-\bar{U}^{\tau}_{i,j})}{8}\bar{U}_{i+1,j}(1-\bar{U}_{i,j})\\
&-\frac{1-p_v(\bar{U}^{\tau}_{i,j}-\bar{U}^{\tau}_{i,j-1})}{8}\bar{U}_{i,j}(1-\bar{U}_{i,j-1})-\frac{1+p_v(\bar{U}^{\tau}_{i,j+1}-\bar{U}^{\tau}_{i,j})}{8}\bar{U}_{i,j}(1-\bar{U}_{i,j+1})\\
&+\frac{1+p_v(\bar{U}^{\tau}_{i,j}-\bar{U}^{\tau}_{i,j-1})}{8}\bar{U}_{i,j-1}(1-\bar{U}_{i,j})+\frac{1-p_v(\bar{U}^{\tau}_{i,j+1}-\bar{U}^{\tau}_{i,j})}{8}\bar{U}_{i,j+1}(1-\bar{U}_{i,j})\\
&+q\bar{U}_{i+1,j}(1-\bar{U}_{i,j})+q\bar{U}_{i-1,j}(1-\bar{U}_{i,j})\\
&+q\bar{U}_{i,j+1}(1-\bar{U}_{i,j})+q\bar{U}_{i,j+1}(1-\bar{U}_{i,j}).
\end{aligned}
\end{equation}
By using the truncated Taylor series, we derive the following continuum limit
\begin{equation}
\label{eq_2D}
\frac{\partial U}{\partial t}=D\frac{\partial}{\partial x}\left[ \frac{\partial U}{\partial x}-h_1(U)\frac{\partial U^\tau}{\partial x}\right]+D\frac{\partial}{\partial y}\left[ \frac{\partial U}{\partial y}-h_2(U)\frac{\partial U^\tau}{\partial y}\right]+f(U),
\end{equation}
where $D=\lim_{\Delta,\eta\to0}\Delta^2/4\eta$ and
\begin{equation}
\label{hf_2D}
h_1(U)=2p_vU(1-U),\quad h_2(U)=2p_hU(1-U), \quad 
f(U)=rU(1-U),
\end{equation}
with $r=\lim_{\eta\to0}4q/\eta$.
Since the derivation steps are similar to the content in Section~\ref{seU_con}, we omit the details. 

\begin{figure}[H]
   \centering \includegraphics[width=\textwidth]{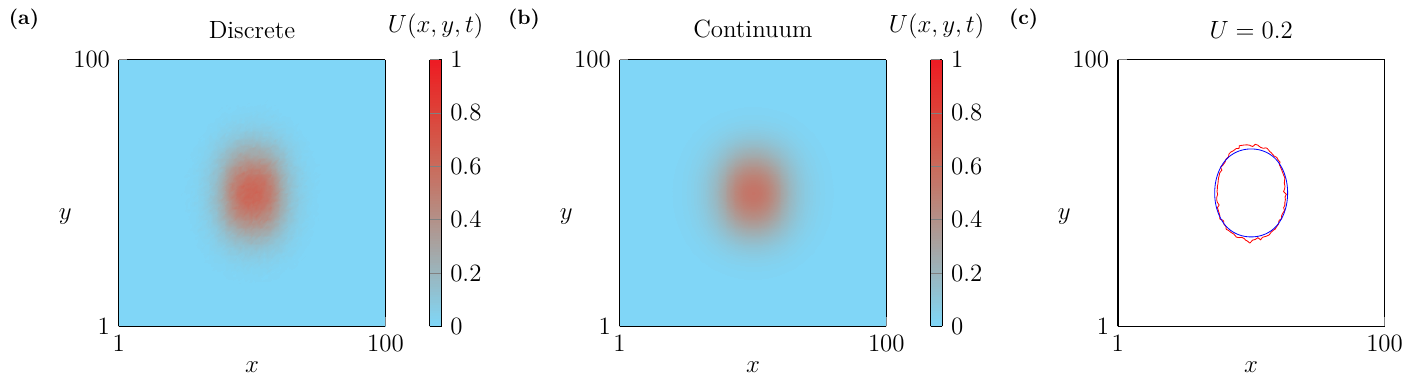}
   \caption{The numerical solutions of \eqref{eq_2} (panel (a)) and the averaged stochastic results (panel (b)). We set $q=1\times10^{-5}$, $p_h=0.8$, $p_v=0$, $\tau=40$, $\Delta=1$ and $\eta=1$, and show the results at $t=200$. The blue and red curve in panel (c) is the density contour of $U(x,y,200)=0.2$ in the numerical solution and the stochastic simulation, respectively. }
   \label{fig8}
\end{figure}

We provide the averaged data and the numerical solutions of~\eqref{eq_2D} in Figure~\ref{fig8}. Here, we consider a $100\times100$ lattice and initially gather the individuals together in the middle of the domain with a $20\times20$ square. For the continuum model, this relates to an initial distribution $U(x,y,0)=1$ in the region $\{(x,y)|x\in[40,60], y\in[40,60]\}$ and $U(x,y,0)=0$ elsewhere. We set $q=1\times10^{-5}$, $p_h=0.8$, $p_v=0$, $\tau=40$, $\Delta=1$,  $\eta=1$ and run the simulation to time $t=400$. Since $p_v=0$ and $p_h>0$, in this case, the memory effect only contributes to the horizontal migration of the population. To compare the discrete and continuum results, we further draw the density contour where $U(x,y,400)=0.2$ in Figure~\ref{fig8}(c). Again, the discrete and continuum results match well. It is obvious from these results that the memory effect influences the propagation dynamics of the population as the propagation is much slower in the horizontal direction than in the vertical direction.

\begin{figure}[H]
   \centering \includegraphics[width=0.95\textwidth]{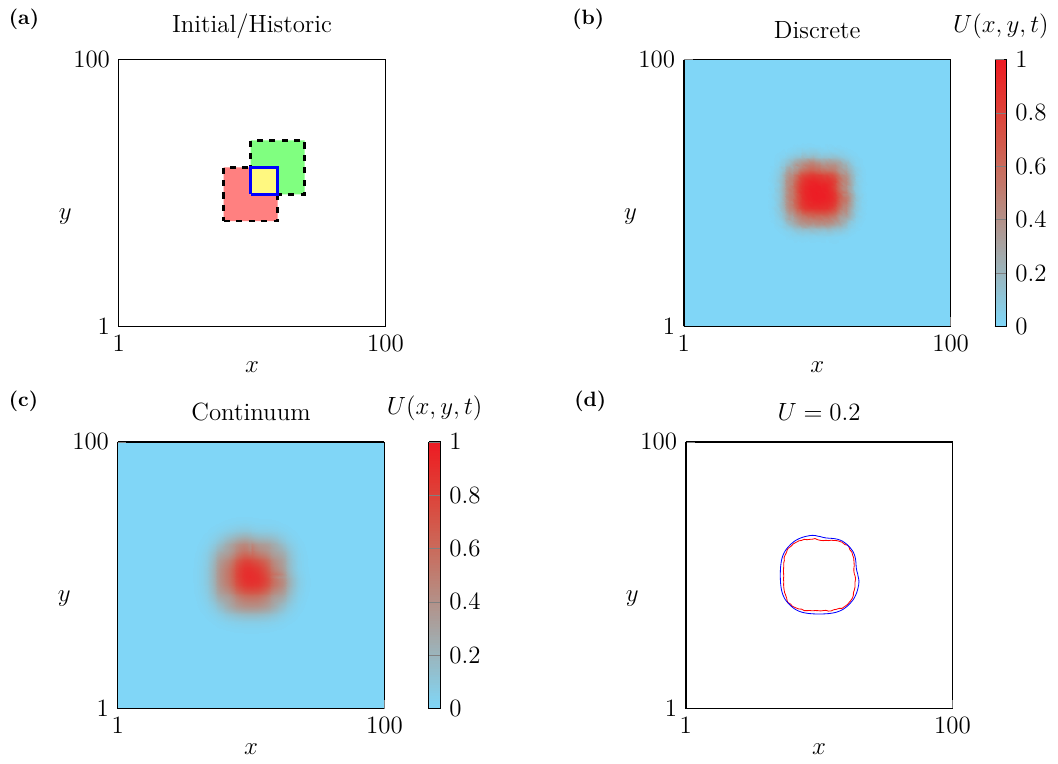}
   \caption{Population dynamics with a historic distribution different from the initial distribution at $t=0$. In panel (a), the historic distribution is painted with green and yellow colours, and the initial distribution at $t=0$ is painted with red and yellow colours. The numerical solution of \eqref{eq_2} and the averaged stochastic result are given in panel (b) and panel (c), respectively. We set $q=1\times10^{-5}$, $p_h=-0.8$, $p_v=-0.8$, $\tau=40$, $\Delta=1$ and $\eta=1$, and show the results at $t=40$. The blue and red curve in panel (d) is the density contour of $U(x,y,200)=0.2$ in the numerical solution and the stochastic simulation, respectively. }
   \label{fig9}
\end{figure}

To provide further insight into our modelling framework we now consider a historic distribution that is different from the initial distribution. As illustrated in Figure~\ref{fig9}(a), the population with $t\in(-\tau,0)$ is placed in the region $\{(x,y)|x\in[50,70], y\in[50,70]\}$. We further set $p_v=p_h=-0.8$ so that the historic distribution equally affects the horizontal and vertical directions. We show the population distribution at time $t=40$ from the discrete model in Figure~\ref{fig9}(b), from the continuum model in Figure~\ref{fig9}(c), and draw the density contour $U(x,y,50)=0.2$ in Figure~\ref{fig9}(d). Although the initial distribution is symmetric about the centre of the domain, the evolving population distribution loses this symmetry owing to the memory effect. However, in this case the memory effect is relatively weak since it only affects the movement of individuals when there is a nonzero density gradient in the distribution at $t\in(-\tau,0)$. For example, during the interval $0 < t < 40$, the memory effect only changes the movement probability of those agents located near the top right corner of the initial distribution as depicted by boundaries highlighted with blue solid lines and black dashed lines in Figure~\ref{fig9}(a).

\section{Conclusion and future work}
In this study, we designed and implemented a new lattice-based model with a focus on understanding how memory-based diffusion relates to the migration properties of individuals within the population. In the population biology modelling literature, the most common way to incorporate the spatial memory effect has been to use a PDE model based on equation~\eqref{eq_shi}. However, an intuitive understanding of why different forms of diffusion and advection terms should be considered has been lacking. We address this knowledge gap by working with an easy-to-interpret discrete model and we derive the corresponding continuum limit PDE model. In the discrete model, migration events follow a random walk process influenced by the crowding effect and the historic distribution of individuals. In the continuum description \eqref{eq_1},
the nonlinear function $h(U)$ intuitively describes the relationship between the memory of individuals and the memory-based diffusion: a negative $p$ suggesting the preference of moving to the historically vacant sites leads to a larger population flux; a positive $p$ suggesting the preference of moving to the historically occupied sites leads to a smaller population flux. This framework supports the modification of memory-based diffusion terms in PDE models, where $h(U)$ associates to a volume-filling type term $U(1-U)$  and $\lvert h(U)\lvert\le1$. The mathematical analysis of the global boundedness of solutions to the resulting continuous model is completed based on a recent theoretical work \cite{Liu2024}. Furthermore, by using this discrete-continuous framework, we interpret how memory effect affects the movement of individuals, leading to $h(U)=2pU(1-U)$ with different choices of $p$ and different speeds of population dispersal.

From the discrete model perspective, the source term $f(U)=rU(1-U)$ is derived following a volume-filling-type argument where the new individuals arising from proliferation events are associated with a certain volume and they cannot overlap or pass through other individuals in the population. Therefore, according to our approximate conservation argument, we modify $f(U)=rU(1-U)$ to $\bar{f}(U)=rU(1-U^k)$ with a simple proliferation dispersion rule: daughter agents are dispersed $k$ lattice sites distant from the mother agent. This extension illustrates how it can be straightforward to generalise the discrete mechanism and then explore how these generalisations impact the continuum-limit PDE description.  We also extend our discrete-continuous modelling framework from a one-dimensional strip region to a two-dimensional plane. In two-dimensional scenarios, the memory effect leads to several interesting population dynamics, including biased migration and irregular dispersal profiles.

There are many ways to extend the work in this study. For example, incorporating the death mechanism of individuals could lead to bistable dynamics, so that the current framework can be extended for exploring the influence of spatial memory effect in the survival or extinction of populations. Although the stochastic algorithm has been extended to higher dimensions, the current model can only reflect the biased movement of individuals horizontally or vertically due to the memory effect. An extension of the current framework for modelling memory effect is integrating the influences along different directions within a uniform structure. Moreover, in this work we consider a very simple memory effect, where the movement of individuals is only influenced by the historic density gradient. However, memory effects are complicated and can be influenced by many factors, such as the path of individuals \cite{FaganEL2013}.
How to incorporate more complicated movement mechanisms associated with memory effects into the discrete-continuous modelling framework is also an interesting question. We leave these extensions for
future consideration.

\section*{Acknowledgments}
Y. Li is supported, in part, by NSFC No.12301624 and China Postdoctoral Science Foundation No.2023M740934. C. Wang is supported by NSFC No.12171118 and Fundamental Research Funds for the Central Universities No.2022FRFK060016.  MJS is supported by the Australian Research Council DP230100025.

\section*{Data availability}
The code used to generate the results presented here can be found on \href{https://github.com/Yifei216/MemoryBased1}{Github}. Data will be made available on reasonable request. 

\bibliography{LiteratureList}

\section*{Appendix}
To numerically solve the memory-based equation
\begin{equation}
\label{PDE_App}
\frac{\partial U}{\partial t}=D\frac{\partial}{\partial x}\left[ \frac{\partial U}{\partial x}-h\left(U\right)\frac{\partial U^\tau}{\partial x}\right]+f(U),
\end{equation}
we consider a uniform discretisation of the domain $0<x<L$, with mesh spacing $\delta x>0$.  Each mesh point is indexed $x_i$ with $i=0,1,2,...,I$ satisfying $I=L/\delta x$.  This approach leads to a system of $I+1$ coupled ordinary differential equations that can be written as 
\begin{equation}
    \nonumber
    \frac{{d}U_{i}}{ {d}t}=
    \dfrac{D}{2\delta x^2}(\bullet)+f(U_{i}),
\end{equation}
where
\begin{equation}
\nonumber
    \bullet
    =U_{i+1}-2U_i+U_{i-1}-\frac{h(U_{i+1})+h(U_i)}{2}(U^\tau_{i+1}-U_i)+\frac{h(U_{i})+h(U_{i-1})}{2}(U^\tau_{i}-U_{i-1}).
\end{equation}
This equation is valid for interior nodes, and can modified on the boundary nodes to model no-flux boundary conditions. The numerical solution is obtained by solving this system of coupled ordinary differential equations using the ``DifferentialEquations.jl'' in Julia for solving delay differential equations~\cite{julia}.
\end{document}